\documentclass[useAMS,usenatbib]{mn2e}

\usepackage{epsfig}
\usepackage{amssymb,latexsym}

\journal{Preprint astro-ph/0308074}
\title[SZ/X-ray cluster scaling relations]
{Hydrodynamical simulations of the Sunyaev--Zel'dovich effect:
cluster scaling relations and X-ray properties}

\author[A.~C.~da Silva et al.]
{Antonio C.~da Silva,$^{1,2}$\thanks{E-mail:antonio.dasilva@ias.u-psud.fr}
 Scott T.~Kay,$^{3}$ Andrew R.~Liddle$^{3}$ and Peter A.~Thomas$^{3}$ \\
$^{1}$IAS, B\^atiment 121, Universit\'e Paris Sud, F-91405, Orsay, France\\
$^{2}$Observatoire Midi-Pyr\'en\'ees, Av. Edouard Belin 14, 31500 Toulouse, 
France \\
$^{3}$Astronomy Centre, University of Sussex, Falmer, Brighton BN1 9QH\\
}

\begin{document}

\maketitle

\label{firstpage}

\begin{abstract}
The Sunyaev--Zel'dovich effect is a powerful new tool for finding and studying 
clusters at 
high redshift, particularly in combination with their X-ray properties. 
In this paper we quantify the expected scaling relations between these 
properties using numerical 
simulations with various models for heating and cooling of the cluster gas. 
For a {\it Non-radiative} model, we find scaling relations in good
agreement with self-similar predictions: $Y \propto T_{\rm X}^{5/2}$
and $Y \propto L_{\rm X}^{5/4}$. Our main results focus on predictions from 
{\it Cooling} and {\it Preheating} simulations, shown by 
Muanwong et al.~(2002) to provide a good match to the X-ray scaling 
relations at $z=0$. For these runs we find slopes of approximately $Y \propto 
T_{\rm X}^{3}$ 
and $Y \propto L_{\rm X}$, steeper and flatter than the self-similar
scalings respectively. We also study the redshift evolution of the
scaling relations and find the slopes show no evidence of evolution out to 
redshifts well beyond one, while the normalizations of relations between 
the SZ signal and X-ray properties do show evolution relative to that 
expected from self-similarity, particularly at $z<1$. 
\end{abstract}


\section{Introduction}

The prospect of surveying for galaxy clusters across a wide range of redshifts 
using the Sunyaev--Zel'dovich (SZ) effect (Sunyaev \& Zel'dovich 1972, 1980) 
raises the hope of a much-improved understanding of cluster physics, and in 
particular of the evolution of the cluster gas. The SZ effect is a 
particularly powerful tool when combined with observations of X-ray emission 
from the hot cluster gas, as the two techniques probe different properties of 
the cluster gas distribution, and with the {\it XMM-Newton} and {\it Chandra} 
X-ray satellites in their main operations phases the number of clusters and 
groups with quality X-ray observations is rapidly increasing. It is therefore 
timely to extend theoretical models of clusters, where efforts have 
hitherto been
concentrated upon their X-ray properties, to incorporate predictions for the 
SZ effect.

Numerical simulations indicate that the distribution of non-baryonic 
dark matter in clusters, which dominates their mass, 
is approximately self-similar (e.g. Navarro, Frenk \& White 1995, 1997).
Observations indicate, however, that the baryonic gas component cannot 
share such a degree of self-similarity. This is particularly evident 
from observations of the $L_{\rm X}$--$T_{\rm X}$ relation, which show it to 
be steeper than predicted by the self-similar model (e.g. 
Edge \& Stewart 1991; Xue \& Wu 2000). 
The reason for this discrepancy is that the gas is 
less centrally concentrated than the dark matter, due to physical processes 
(in addition to gravity) that raised the entropy of the gas (e.g. 
Evrard \& Henry 1991; Kaiser 1991;
Bower 1997; Voit et al. 2002; Ponman, Sanderson \& Finoguenov 2003). 
This entropy could come from direct heating of the hot gas 
from stars or AGN (e.g. Wu, Fabian \& Nulsen 2000; Bower et al. 2001; 
Quilis, Bower \& Balogh 2001), 
or from radiative cooling which transforms low-entropy material into 
stars allowing high-entropy material to flow in to replace it 
(e.g. Knight \& Ponman 1997; Pearce et al. 2000; Bryan 2000). 
Understanding the relative importance of 
heating and cooling, and how these processes conspire to establish excess
entropy in clusters, is one of the key problems in cluster gas physics.

The implications of an excess of entropy on the global properties of the SZ 
effect (e.g. source counts, the mean Compton parameter and the 
angular power spectrum) has previously been investigated  
using both analytical techniques (e.g. Cavaliere \& Menci 2001; 
Holder \& Carlstrom 2001; Zhang \& Wu 2003) and numerical simulations 
(e.g. Springel, White \& Hernquist 2001; da Silva et al. 2001; 
White, Hernquist \& Springel 2002). The mass dependence of the
SZ effect in clusters has been addressed by Metzler (1998) and by 
White et al.~(2002). More recently, McCarthy et al. (2003a) used a 
semi-analytic model 
to derive scaling relations between the central 
Compton parameter and mass, temperature, SZ flux density and 
X-ray luminosity. These scalings were subsequently used in a companion 
paper (McCarthy et al. 2003b) to estimate the level of the 
entropy floor compatible with present-day SZ observations.

The purpose of this paper is to investigate the relationship between
SZ and X-ray properties of clusters in $N$-body/hydrodynamical 
simulations that include models for both cooling and preheating.
The simulations we use have already been demonstrated to give 
good overall agreement with
observed X-ray scaling relations at redshift zero (Muanwong et al.
2001; Thomas et al. 2002; Muanwong et al. 2002).
Our approach is mainly focused on the derivation of theoretical 
scaling laws, and their evolution with redshift, between 
the integrated SZ flux density and other cluster properties, such
as X-ray temperature and luminosity. We quantify deviations 
from self-similar evolution on both the slope and normalization of the 
scalings and provide fits which can be used for comparison with 
other theoretical models and observations. An observationally-motivated
analysis of SZ/X-ray correlations, that can be more readily applied to `blind' 
searches for high-redshift clusters, will be presented in a future paper.

This paper is organized as follows. We define the observational 
quantities used and the simple scaling laws
predicted by the self-similar model in Section~2. Details of our
simulations and how our cluster catalogues were constructed are 
presented in Section~3. In Section~4, we investigate the correlation 
between thermal SZ integrated flux and other 3D cluster properties 
(mass, mass-weighted and emission-weighted temperature and X-ray 
luminosity) from the simulation data at redshift zero. The evolution of 
these relations with redshift is studied in Section~5, before we draw 
conclusions in Section~6.

\section{Theoretical Framework}

\subsection{Definitions of physical quantities}

The SZ effect arises due to inverse Compton scattering
of CMB photons off free electrons, with the largest SZ signal 
being produced by intracluster gas. The total SZ flux density 
produced by a cluster is the integral of the SZ sky brightness 
over its subtended solid angle 
%
\begin{equation}
S_{\nu} = I_0 \,\int {\left[ {g(x)\,y-h(x)\,
b }\right] \,d\Omega }\, , 
\label{eq_1}
\end{equation}
where $I_0=2\left( {k_{\rm B}T_0} \right)^3/(hc)^2 \simeq 2.28\times 10^4$~mJy
arcmin$^{-2}$, 
\footnote{1 {\rm mJy} = $10^{-26}$ erg s$^{-1}$ m$^{-2}$ Hz$^{-1}$.} 
for a mean CMB temperature of $T_0 \simeq 2.725$~K (Mather et al. 1999).
The first term inside the brackets accounts for the thermal SZ effect, 
due to internal motion of the electrons, whereas the second term, 
the kinetic SZ effect,
is the contribution due to bulk motion of the gas. We ignore the kinetic
SZ effect in this paper as the thermal SZ effect dominates at all 
frequencies except $\nu\simeq 217$GHz ($x=h\nu /k_{\rm B}T_0\simeq 3.83$).

The Comptonization parameter $y$ contains information on the
structure of the intracluster gas
%
\begin{equation}
y={{k_{\rm B}\sigma _{\rm T}} \over {m_{\rm e}c^2}}
\int {T_{\rm e}\,n_{\rm e}\,dl},
\label{eq_2}
\end{equation}
where $T_{{\rm e}}$ and $n_{{\rm e}}$ are the temperature
and density of the electrons, $\sigma_{{\rm T}} \simeq 6.65\times 10^{-25}\,
{\rm cm}^2$ the Thomson cross-section, $c$ the speed of light and
$m_{{\rm e}}$ the electron rest mass. It is therefore convenient to 
define the integrated contribution to $S_{\nu}$ from $y$
%
\begin{equation}
Y = \int y \,\,d\Omega
=\int {\frac{y}{d_A^2}}\,dA
={ {k_B\sigma _{\rm T}} \over {m_{\rm e}c^2\,d_A^2}} 
\int _{V} T_{{\rm e}} n_{{\rm e}} \,dV,
\label{eq_3}
\end{equation}
where $d_A$ is the angular diameter distance 
and $dA=d\Omega \,d_A^2$ is the sky projected area element of the cluster.
Since $Y$ depends on distance, we will usually work 
with the intrinsic thermal SZ signal, defined as 
$Y^{\rm int}=Y\,d_A^2(z)$.

The X-ray luminosity of a cluster is 
\begin{equation}
L_{\rm X} =\int \, n_{\rm e} n_{\rm H}
\Lambda_{\rm c}(T)\, dV.
\label{eq_4}
\end{equation}
The integral is over the volume of the cluster and
$\Lambda_{\rm c}(T)$ denotes the cooling rate of the gas,
assuming collisional ionization equilibrium. At temperatures
above a few keV, the emission is dominated by thermal bremsstrahlung,
but below this line emission becomes important.

\subsection{Self-similar scaling relations}

In the absence of any non-gravitational heating and cooling
processes, clusters scale self-similarly to a good approximation
(Kaiser 1986; Navarro et al. 1995). In self-similar models, the
temperature of the gas scales with the cluster mass as
\begin{equation}
T \propto \, M^{2/3} (1+z)\,,
\label{eq_5}
\end{equation}
assuming the density of the gas scales with the mean density  
$\rho \propto (1+z)^3$ and the system is in virial equilibrium.
Applying this to Eq.~(\ref{eq_3}) we get
%
\begin{equation}
Y^{\rm int} \propto \,
\left\{ \begin{array}{ll}
	f_{\rm gas}\, T^{5/2}\, (1+z)^{-3/2}\,\\
	f_{\rm gas}\, M^{5/3}\, (1+z)\,,\\
\end{array}
\right.	
\label{eq_6}
\end{equation}
where $f_{\rm gas}$ is the gas mass fraction of the cluster. 

The X-ray luminosity of a self-similar cluster scales as
%
\begin{equation}
L_{\rm X} \propto \,
\left\{ \begin{array}{ll}
	f^2_{\rm gas}\, T^{2}\, (1+z)^{3/2} \, \\
	f^2_{\rm gas}\, M^{4/3} \, (1+z)^{7/2} \,,\\
\end{array}	
\right.
\label{eq_7}
\end{equation}
assuming that the X-rays are due to bremsstrahlung emission 
($\Lambda_{\rm c}\propto T^{1/2}$).

\section{Method}

\subsection{Simulation details}

Results are presented from three simulations, described
in detail elsewhere (Thomas et al.~2002; Muanwong et al.~2002); 
we summarize pertinent details here.

We assume a $\Lambda$CDM cosmology, setting the 
density parameter, $\Omega_{{\rm m}}=0.35$, cosmological 
constant, $\Omega_{\Lambda}=0.65$, hubble parameter, 
$h=0.71$, baryon density, $\Omega_b h^2=0.019$, CDM power
spectrum shape parameter, $\Gamma=0.21$, and normalization 
$\sigma_8=0.9$. These values are similar to those favoured by current 
observations including {\it WMAP}.
Initial conditions were generated using
$160^3$ particles each of baryonic and dark matter, perturbed 
from a regular grid of comoving length, $L=100 h^{-1} {\rm Mpc}$. 
These choices set the gas and dark matter particle masses equal to 
$2.6 \times 10^{9} h^{-1} {\rm M_{\odot}}$ 
and  $2.1 \times 10^{10} h^{-1} {\rm M_{\odot}}$ respectively. 

The runs were started at redshift $z=49$ and evolved to $z=0$ using a
parallel version of the {\sc hydra} $N$-body/hydrodynamics code 
(Couchman, Thomas \& Pearce 1995; Pearce \& Couchman 1997), 
which uses a Smoothed Particle Hydrodynamical (SPH) implementation
similar to that discussed in Thacker \& Couchman (2000), and conserves
both energy and entropy to a satisfactory degree.
The gravitational softening was fixed at 
$50\,h^{-1} {\rm kpc}$ in comoving co-ordinates until $z=1$, then
at $25\,h^{-1} {\rm kpc}$ in physical co-ordinates until $z=0$.

The simulations differed only in the adopted heating/cooling model.
In the first model, a {\it Non-radiative} simulation, the gas was
subject only to gravitational, adiabatic and viscous forces, 
and so could only change its entropy through shock-heating.
This model fails to reproduce the X-ray scaling
relations (Muanwong et al.~2001) but nevertheless can be used
to test the self-similar scalings presented in the
previous section.

The second simulation, a {\it Cooling} model, allowed gas to
cool radiatively as described by Thomas \& Couchman (1992).
Tabulated values of $\Lambda_{\rm c}$ were generated from cooling
tables in Sutherland \& Dopita (1993), assuming a time-varying global 
gas metallicity, $Z=0.3(t/t_0) Z_{\odot}$, where 
$t_0 \simeq 12.8 \, {\rm Gyr}$ is the current age of the Universe.
Cooled material (identified as all gas particles 
with temperatures $T < 1.2 \times 10^4$ K and overdensities $\delta > 1000$) 
was converted into collisionless particles. Note that the choice of density 
threshold (considerably lower than the observed density of star formation)  
largely circumvents the problem of unphysical cooling of 
hot gas particles in contact with cold dense gas, as well as increasing 
run-time efficiency. 
The global fraction of 
cooled gas in the simulation box is $\sim$15 per cent at $z=0$. The
removal of this gas from the hot phase sufficiently increases the entropy
in the centres of clusters to reproduce the X-ray scaling relations,
but the cooled fraction is 2 to 3 times higher than observed 
(Cole et al.~2001). 
In this respect, this run provides a maximum variation of $Y$ 
from a cluster due to effect of radiative cooling alone. 

Finally, the third simulation, a {\it Preheating} model, also
allowed the gas to cool radiatively, but the gas was impulsively
heated by 1.5 keV per particle at $z=4$. This generates the 
required minimum entropy in the centres of clusters but prevents further 
cooling to $10^4$K; the global fraction of cooled gas in the 
simulation box is only $\sim$0.5 per cent at $z=0$.

Note that we do not attempt to perform a numerical resolution study
for this work. Apart from being computationally demanding, it would be
misleading to study the effects of resolution on our current
models. In particular, increasing the resolution in the {\it Cooling}
model would result in an even higher cooled fraction, and so attaining
convergence in this model is undesirable.  Instead we interpret our
models as a relatively crude first attempt at capturing the effects of
increasing the entropy of the gas by the required amount in two
completely different ways (i.e.~from cooling and heating). A proper
convergence analysis is warranted when feedback is included
(e.g.~Springel \& Hernquist 2003) and will be investigated in future
work.

\subsection{Cluster selection and SZ/X-ray estimators}

We identified clusters in our simulations according to the method 
described in Thomas et al (1998).
Clusters were selected by first creating a minimal-spanning
tree of all dark matter particles whose density exceeds 
$\delta = 178 \Omega^{-0.55}(z)$ times the mean dark matter density (the 
approximate value for a virialized sphere, as predicted by the 
spherical top-hat model; Eke, Navarro \& Frenk 1998). 
The tree was then pruned into clumps 
using a maximum linking length equal to $0.5\delta^{-1/3}$ times
the mean interparticle separation. A sphere was then grown 
around the densest particle in each clump until the enclosed mean density 
exceeded a value $\Delta$ in units of the comoving critical density.
For this paper, we use $\Delta=200$, larger than the virial value 
for our cosmology ($\Delta \sim 111$) but commonly used by other
authors. Master cluster catalogues\footnote{The catalogues are available to 
download at the website {\tt virgo.sussex.ac.uk}} were produced containing
only objects with at least 500 particles each of gas and dark matter,
equivalent to a mass limit, 
$M_{\rm lim}\approx1.18\times10^{13}h^{-1}M\sun $. 
At $z=0$, the catalogues contain 428, 457 and 405 clusters 
for the {\it Non-radiative}, {\it Cooling} and {\it Preheating} simulations 
respectively.

For each cluster, we then estimated various observable quantities:
the intrinsic SZ signal, $Y^{\rm int}$, the X-ray emission-weighted
gas temperature, $T_{\rm X}$ and the X-ray luminosity, $L_{\rm X}$.
We identified the hot intracluster gas as all gas particles within
$R_{200}$ with $T>10^5$K (in practice there are very few particles with 
$10^4$K$<T<10^5$K as their cooling times are very short) and assume full 
ionization, such that the number of electrons per baryon is $\eta=0.88$ for a 
hydrogen mass fraction $X=0.76$. We have
\begin{equation}
Y^{\rm int} = 
\frac{k_{\rm B}\sigma_{\rm T}}{m_{\rm e}c^2}\frac{\eta}{m_{\rm p}}
\, f_{\rm gas} \, M \, T_{\rm mw}, 
\label{eq_8}
\end{equation}
where $T_{\rm mw} = {\sum \, m_i \, T_i/ \sum \, m_i }$ is the mass-weighted 
temperature of the gas, $f_{\rm gas} = \sum \, m_i /M$ the gas fraction,
$m_i$ the mass of hot gas particle $i$ and $M$ the total mass of the 
cluster. 

The X-ray emission-weighted temperature was estimated as
%
\begin{equation}
T_{\rm X} = \frac{\Sigma_i m_i\rho_i\Lambda_{\rm soft}(T_i,Z)T_i}
              {\Sigma_i m_i\rho_i\Lambda_{\rm soft}(T_i,Z)},
\label{eq_9}
\end{equation}
where $\rho_i$ and $T_i$ are the density and temperature
of the hot gas particles. Since most observed temperatures
use instruments sensitive to soft X-rays, we use a soft-band
cooling function, $\Lambda_{\rm soft}$, from Raymond \& Smith (1977) 
for an energy range 0.3--1.5\,keV. 

The bolometric X-ray luminosity of each cluster, corrected from 
its soft-band emission, was estimated as
%
\begin{equation}
L_{\rm X}={\Lambda_{\rm bol}(T_{\rm X})\over\Lambda_{\rm soft}(T_{\rm X})}
\sum_i{m_i\rho_i\Lambda_{\rm soft}(T_i,Z)\over(\mu m_{\rm H})^2},
\label{eq_10}
\end{equation}
where $\mu m_{\rm H}=10^{-24}$g is the mean molecular mass of
the gas and $\Lambda_{\rm bol}$ is the 
bolometric cooling function (Sutherland \& Dopita 1993). 

Finally, to avoid introducing artificial trends in the derived scalings with
temperature and X-ray luminosity (Sections 4 \& 5), we imposed
lower limits on $T_{\rm X}$ and $L_{\rm X}$ (by inspecting the $T_{\rm
X}$--$M$ and $L_{\rm X}$--$M$ relations at all redshifts) to create
(separate) catalogues that were complete in the two quantities. At
each redshift, these limits approximately correspond to the maximum
temperature and luminosity of all clusters at our mass limit, $M_{\rm lim}$. 
For example, when investigating temperature scalings at $z=0$,
all clusters with $T_{\rm X}< 0.35, 0.65, 0.60$ keV 
(for the {\it Non-radiative}, {\it Cooling} and {\it Preheating} simulations 
respectively) were discarded, reducing the number of clusters in our 
catalogues by around 30 per cent in the {\it Cooling} and {\it Preheating} 
simulations and 46 per cent in the {\it Non-radiative} run.
Similarly, for scalings with X-ray luminosity at $z=0$,
our original cluster 
catalogues were trimmed by selecting only clusters with 
$L_{\rm X} > 2.9 \times 10^{43}$, $9.0\times 10^{41}$, 
$4.5\times 10^{41}\, {\rm erg} \; {\rm s}^{-1}$, for the {\it Non-radiative},
{\it Cooling} and {\it Preheating} simulations, reducing the number of 
clusters in the original catalogues by about 52, 46, and 29 per cent 
respectively.

\section{Scaling relations at redshift zero}

\subsection{The $Y$--$M$ and $Y$--$T_{\rm mw}$ relations}

\begin{figure}
\centering \leavevmode\epsfysize=5.8cm \epsfbox{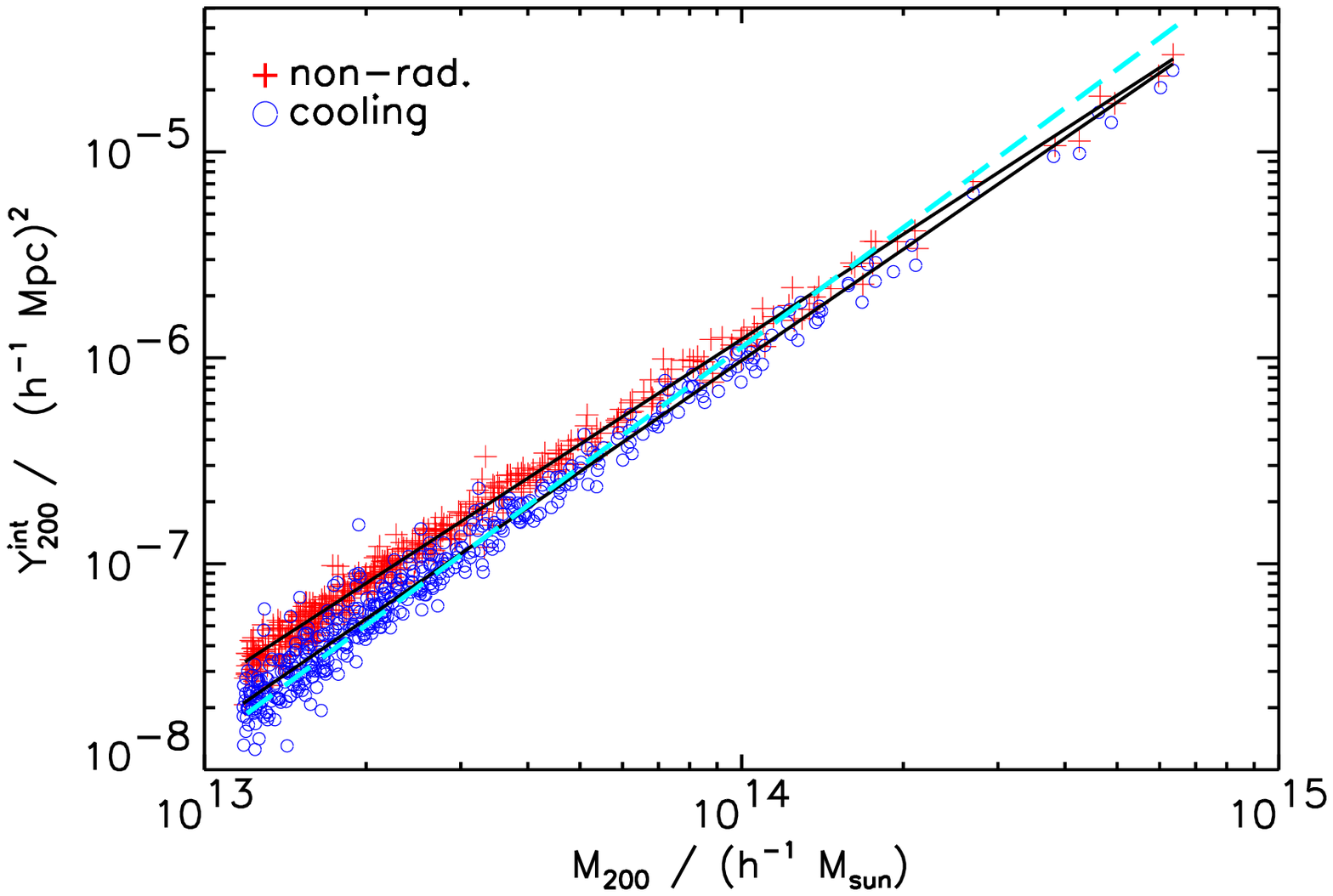}\\ 
\centering \leavevmode\epsfysize=5.8cm \epsfbox{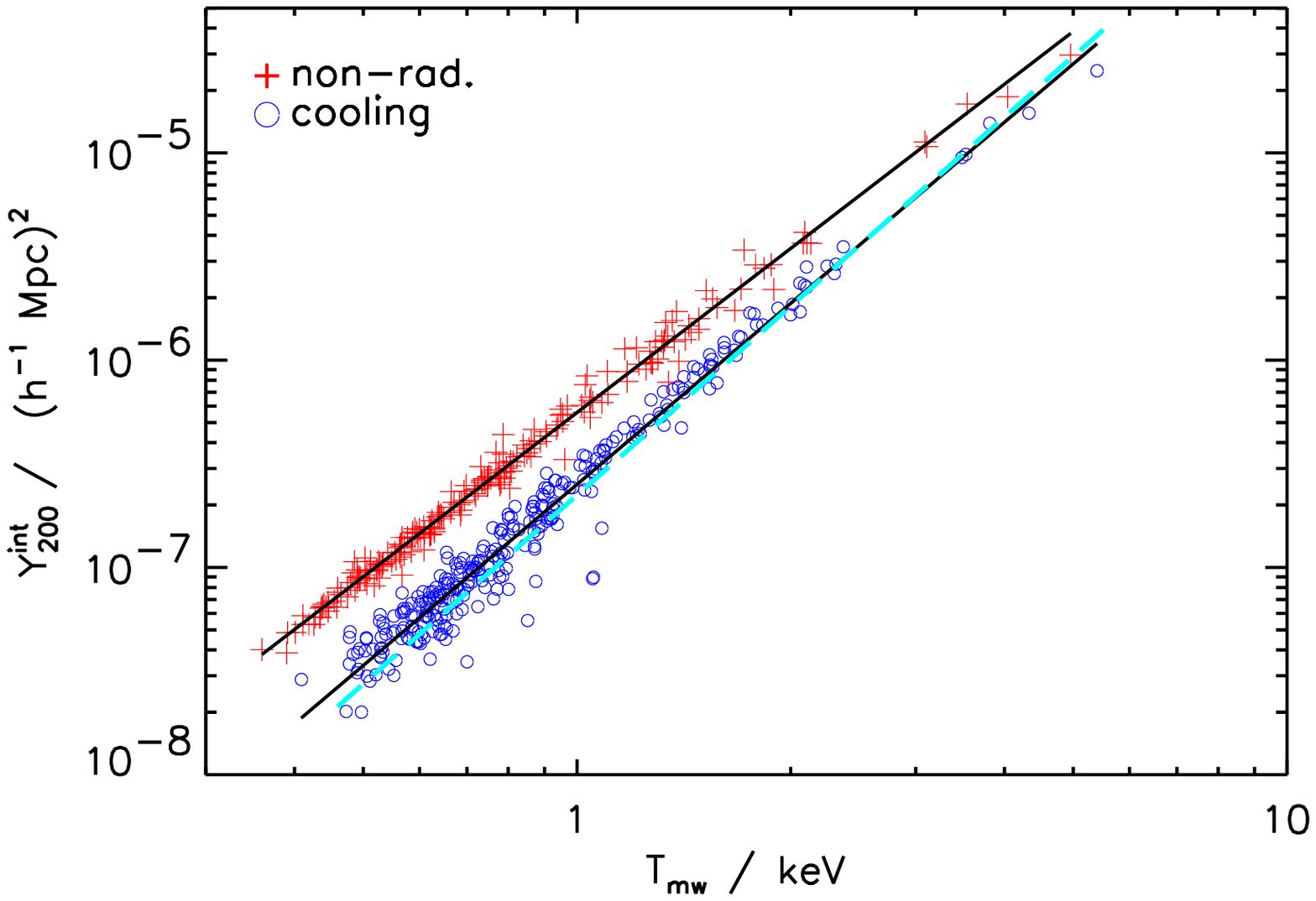}\\
\caption[fig1]
{Scaling relations between $Y^{\rm int}_{200}$ \& $M_{200}$ 
(top panel) and $Y^{\rm int}_{200}$ \& $T_{\rm mw}$ (bottom panel)
for the {\it Non-radiative} (crosses) and {\it Cooling} (circles) simulations. 
Solid lines are power-law fits to the data from these runs,
while the dashed line is the corresponding fit for clusters in
the {\it Preheating} simulation.}
\label{fig:ymt}
\end{figure}

We begin by correlating $Y^{\rm int}$ with mass and mass-weighted
temperature, all measured within $R_{200}$. An early study 
of the relation between mass and SZ absorption in clusters was 
made in unpublished work by Metzler (1998), who verified with non-radiative 
simulations 
the related self-similar scaling, $y \propto M$. More recently, 
White et al. (2002) also recovered the self-similar scaling,
$Y \propto M^{5/3}$, from simulations  
with the same box size but a slightly higher (factor of 2.5) mass resolution
than that used here, concluding that the combined effect of cooling
and feedback (galactic winds) on the SZ properties of their clusters 
was small. 

Fig.~\ref{fig:ymt} illustrates our $Y^{\rm int}_{200}-M_{200}$ and 
$Y-T_{\rm mw}$ relations 
for clusters in the {\it Non-radiative} (crosses) and {\it Cooling} (circles) 
simulations. The solid lines represent power-law best fits to the data, while 
the dashed line is the best-fit to clusters in the {\it Preheating} 
simulation, omitted for the purpose of clarity.
The figure shows that $Y^{\rm int}_{200}$ is tightly correlated with 
mass and mass-weighted temperature in all runs.
As we shall see, however, these correlations show significantly less scatter 
than those relating $Y$ with X-ray emission-weighted properties, because the 
latter is sensitive to substructure in the dense core gas.
The best power-law fits to the $Y^{\rm int}_{200}-M_{200}$ relation in each 
simulation are
\begin{itemize}
\item {\it Non-radiative run}:
\begin{equation} 
Y^{\rm int}_{200}=1.22\times 10^{-6}\,\left( {
\frac{M_{200}}{10^{14}\,h^{-1}\,{\rm M}_\odot} } \right)^{1.69}\,
\left( { h^{-1}\,{\rm Mpc} } \right)^2 ,
\label{cp6_19}
\end{equation}
\item {\it Cooling run}:
\begin{equation}
Y^{\rm int}_{200}=0.97\times 10^{-6}\,\left( {
\frac{M_{200}}{10^{14}\,h^{-1}\,{\rm M}_\odot} } \right)^{1.79}\,
\left( { h^{-1}\,{\rm Mpc} } \right)^2 ,
\label{cp6_20}
\end{equation}
\item {\it Preheating run}:
\begin{equation}
Y^{\rm int}_{200}=1.12\times 10^{-6}\,\left( {
\frac{M_{200}}{10^{14}\,h^{-1}\,{\rm M}_\odot} } \right)^{1.93}\,
\left( { h^{-1}\,{\rm Mpc} } \right)^2 \,.
\label{cp6_21}
\end{equation}
\end{itemize}

\begin{figure}
\centering \leavevmode\epsfysize=5.8cm \epsfbox{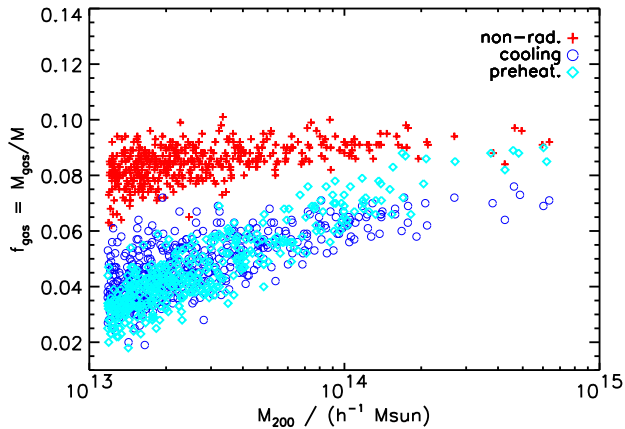}
\caption[Fraction of hot gas within $R_{200}$ as a function of
mass for clusters in the
{\it Non-radiative}, {\it Cooling} and {\it Preheating} 
simulations.]{\label{fig:fhgas} 
Fraction of hot gas within $R_{200}$ as a function of mass
for clusters in the {\it Non-radiative}
(crosses), {\it Cooling} (circles) and {\it Preheating} (diamonds)
simulations. }
\end{figure}

As expected, the slope of the fit to the {\it Non-radiative} 
clusters is close to 5/3, the self-similar value
given in Eq.~(\ref{eq_6}). The slope is steeper for the 
{\it Cooling} and {\it Preheating} simulations however, primarily due 
to lower $Y$ values for low-mass systems. This difference
is due to the effects of cooling and heating on both the
gas fraction and temperature of the clusters.
Cooling removes low-entropy (core) gas from the halo, 
causing higher-entropy material to flow in to replace it. This results in a 
lower gas fraction although
the remaining gas is hotter (Pearce et al.~2000; Bryan 2000; 
Muanwong et al.~2001). For low-mass clusters 
($M_{200}<10^{14}h^{-1}\,$M$_{\odot}$), the net effect of cooling
on $Y$ is dominated by the decrease in hot gas fraction, shown in 
Fig.~\ref{fig:fhgas}, around a factor of two lower than the
{\it Non-radiative} average at $M_{200}=10^{13}h^{-1}\,$M$_{\odot}$. 
For the highest mass clusters this difference decreases to around
$30$ per cent; combined with the increase in temperature 
of the gas, the clusters have very similar $Y$ values in the
two simulations.

In the {\it Preheating} case, the energy injection at high redshift heats 
the gas and expels some of it from the cluster. Hence its effect on the 
cluster properties is similar to the cooling model but the fate of the gas 
is different (Muanwong et al.~2002). Fig.~\ref{fig:fhgas} shows that the gas 
fraction in {\it Preheating} clusters increases more rapidly with mass than for
objects in the {\it Cooling} simulation
(please refer to Section 3.1 and Fig.~3 of Muanwong et al. 2002 for
a detailed discussion of the hot, cold and total baryonic mass
fractions in clusters for the same set of simulations).
In low-mass systems the heating energy
was sufficient to significantly decrease $f_{\rm gas}$ (and hence $Y$),
but left gas fractions almost unchanged in the highest-mass clusters, where
the heating resulted only in a modest increase in temperature. Note,
therefore, that the best-fit relation is only a good description of the
low-mass clusters in the {\it Preheating} simulation; systems more massive
than $\sim 10^{14} h^{-1} M_{\odot}$ are adequately described by the 
{\it Non-radiative} relation.
This also explains why White et al. (2002) find good agreement with the 
self-similar $Y_{\rm int}-M$ scaling, as they only investigated high-mass 
clusters ($M_{200} \ge 5\times 10^{14} h^{-1} {\rm M_{\odot}}$), where
a high fraction of the wind material would have been retained.

\begin{figure}
\centering \leavevmode\epsfysize=5.8cm \epsfbox{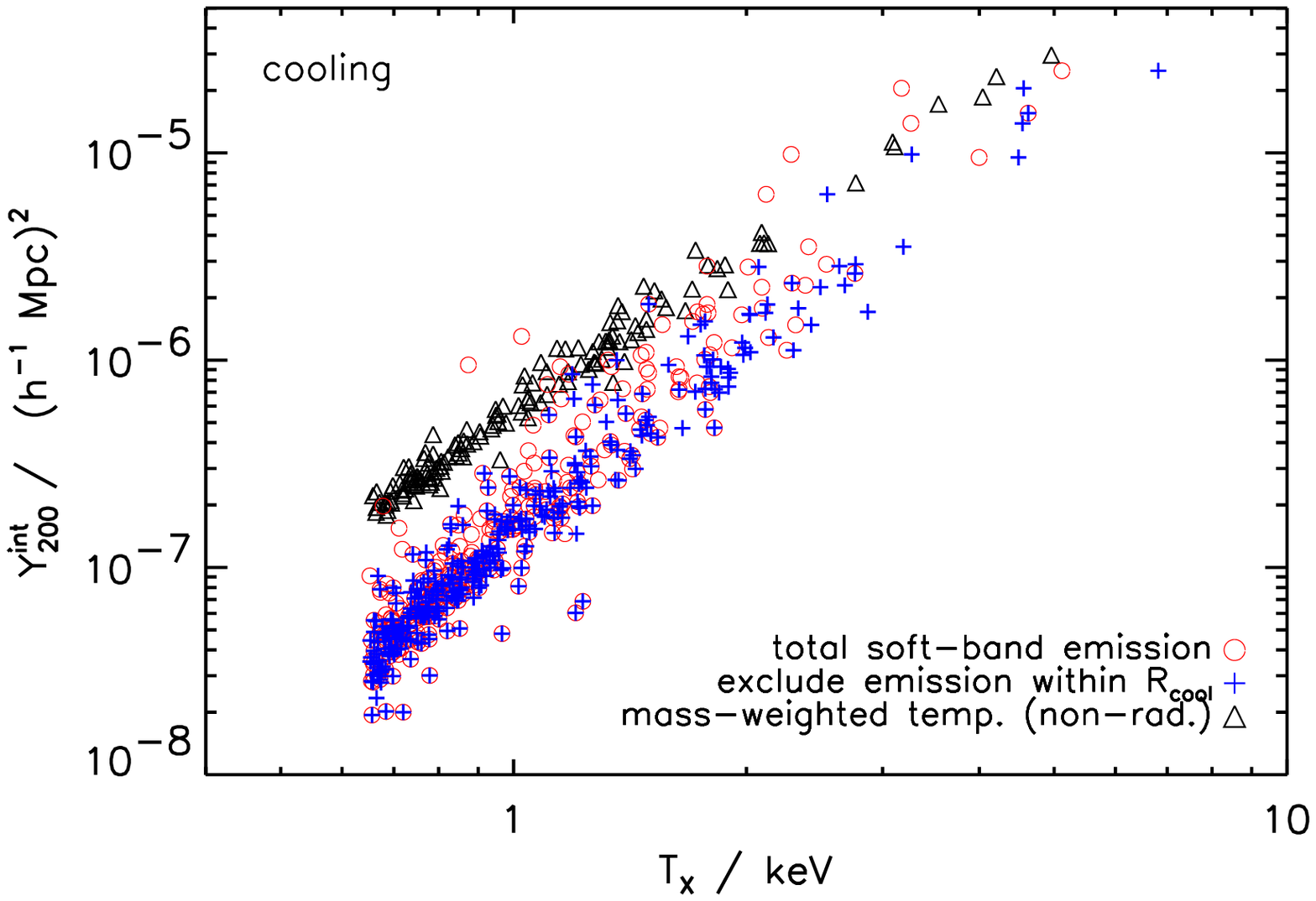}\\ 
\centering \leavevmode\epsfysize=5.8cm \epsfbox{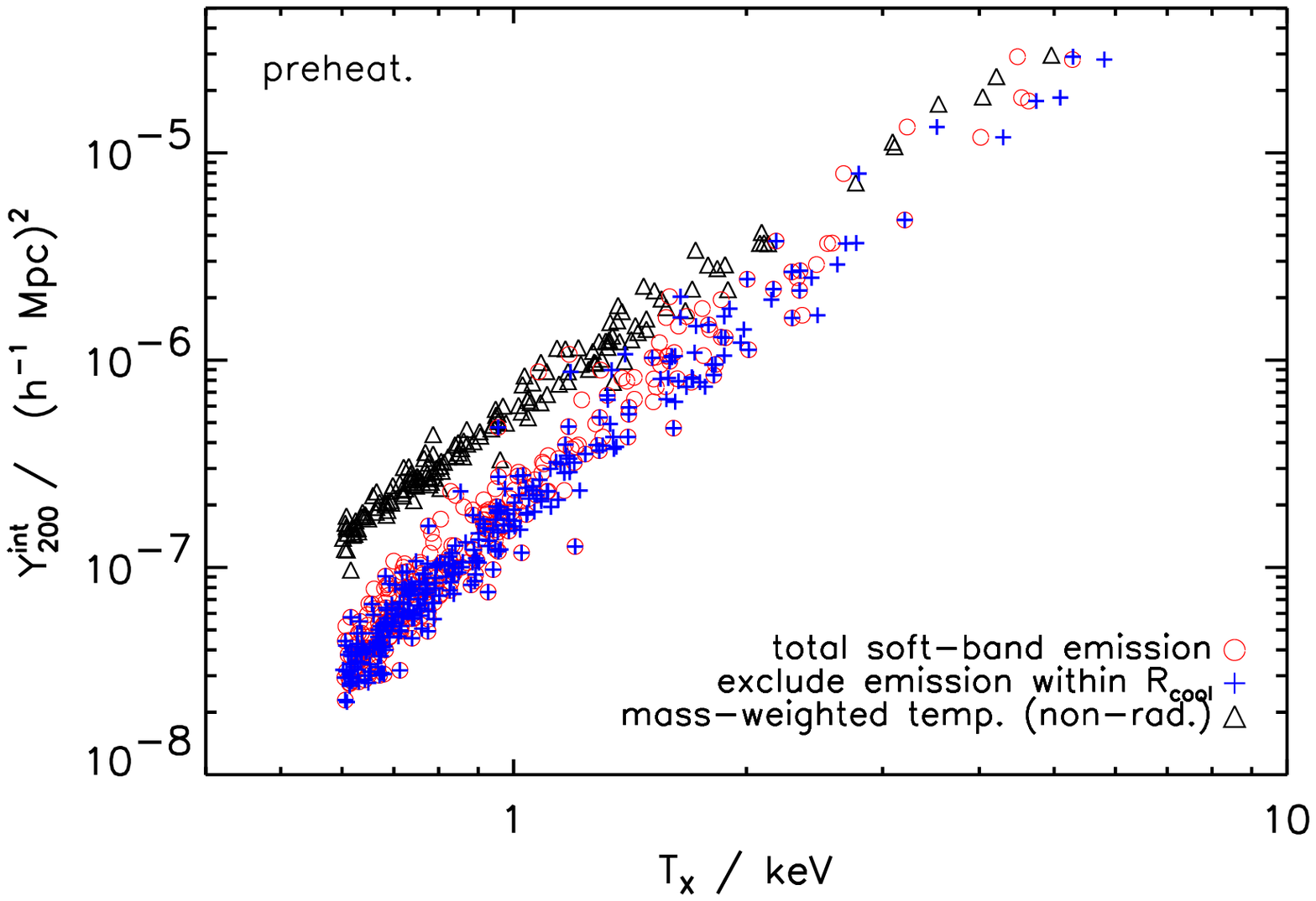}\\
\caption[ ]
{$Y^{\rm int}_{200}$ versus $T_{\rm X}$ for clusters in the {\it Cooling} 
(top panel) and {\it Preheating} (bottom panel) simulations, including
(circles) and excluding (crosses) soft-band emission from within
the cooling radius. Also shown for comparison is the scaling with the 
mass-weighted temperature found in the {\it Non-radiative} simulation 
(triangles).}
\label{fig:ytx}
\end{figure}

The effects of cooling and preheating on the cluster properties
are more readily apparent in the relation between $Y$ and  
mass-weighted temperature. For a cluster of fixed gas and total 
mass, $Y \propto T_{\rm mw}$, and so heating the gas causes a shift
to the right of the steeper self-similar relation ($Y \propto T^{5/2}$). 
Additionally, the decrease in gas fraction lowers $Y$, separating the 
two relations further.
The best-fit $Y^{\rm int}_{200}-T_{\rm mw}$ relations are
\begin{itemize}
\item {\it Non-radiative run}:
\begin{equation} 
Y^{\rm int}_{200}=5.59\times 10^{-7}\,\left( {
\frac{k_{\rm B}T_{\rm mw}}{1\,{\rm keV}} } \right)^{2.63}\,
\left( { h^{-1}\,{\rm Mpc} } \right)^2 
\label{cp6_22}
\end{equation}
\item {\it Cooling run}:
\begin{equation}
Y^{\rm int}_{200}=2.51\times 10^{-7}\,\left( {
\frac{k_{\rm B}T_{\rm mw}}{1\,{\rm keV}} } \right)^{2.90}\,
\left( { h^{-1}\,{\rm Mpc} } \right)^2 
\label{cp6_23}
\end{equation}
\item {\it Preheating run}:
\begin{equation}
Y^{\rm int}_{200}=2.23\times 10^{-7}\,\left( {
\frac{k_{\rm B}T_{\rm mw}}{1\,{\rm keV}} } \right)^{3.03}\,
\left( { h^{-1}\,{\rm Mpc} } \right)^2\, .
\label{cp6_24}
\end{equation}
\end{itemize}

\subsection{The $Y-T_{\rm X}$ relation}

Of more practical interest is to investigate how $Y$ correlates
with X-ray observables, particularly luminosity and emission-weighted
temperature. 

Fig.~\ref{fig:ytx} illustrates the relationship between $Y^{\rm int}_{200}$
and $T_{\rm X}$ for the {\it Cooling} (top panel) and {\it Preheating} (bottom 
panel) simulations.
Circles represent clusters for which all hot gas particles within $R_{200}$
are considered, whereas crosses are for the same clusters but excluding 
particles within the cooling radius in the computation of the temperature. 
We define the cooling radius as the radius within which the gas 
has a mean cooling time of 6 Gyr. (This procedure is an attempt to 
provide a `cooling flow' correction to the estimated values of $T_{\rm X}$.)

Comparing these results with Fig.~\ref{fig:ymt}, we see that 
the scatter is larger for the correlation with $T_{\rm X}$ than with
$T_{\rm mw}$. This is because $T_{\rm X}$ is weighted by density and 
temperature and is therefore more sensitive to substructure,  
particularly from the centres of clusters. This is also evident from 
the fact that the scatter is smaller when the gas within the cooling
radius is excluded (note that this also increases $T_{\rm X}$ in
high-mass systems, which slightly flattens the $Y-T_{\rm X}$ relation).

Table~\ref{tab_1} lists best-fit coefficients, $\alpha_{\rm T}$ and 
$A_{\rm T}$, to the power-law, $Y^{\rm int}_{200}=(A_{\rm T}/10^7)\,
(k_{\rm B}T_{\rm X}/1{\rm keV})^{\alpha _{\rm T}}$
for the {\it Cooling} and {\it Preheating} runs. 
(We also present results from the {\it Non-Radiative} run, using 
a bolometric cooling function, and this agrees well with the self-similar
scaling $Y \propto T^{5/2}$.)
For a given run, the normalization of the $Y^{\rm int}_{200}-T_{\rm X}$ 
relation is lower and slope steeper than for the 
$Y^{\rm int}_{200}-T_{\rm mw}$ relation. 
The heavier weighting of the denser gas exacerbates the effects of 
cooling/heating on the $Y-T$ relation discussed in the previous section.

\begin{table}
\caption{Power-law fits to the simulated cluster $Y-T_{\rm X}$ 
and $Y-L_{\rm X}$ scalings at $z=0$. Here $\alpha_{\rm T}$ and 
$\alpha_{\rm L}$ are the slopes of the $Y^{\rm int}_{200}-T_{\rm X}$ and 
$Y^{\rm int}_{200}-L_{\rm X}$ 
relations, whereas $A_{\rm T}/10^{7}$ and $A_{\rm L}/10^{7}$ are the values 
of  $Y^{\rm int}_{200}$ derived from these relations at  
$k_{\rm B}T_{\rm X}=1$keV and 
$L_{\rm X}=10^{43}h^{-2}\, {\rm erg} \, {\rm s}^{-1}$, respectively.
}
\label{tab_1}
\begin{tabular}{llcccc}
\hline
         && \multicolumn{2}{c}{$Y^{\rm int}-T_{{\rm X}}$}&
         \multicolumn{2}{c}{$Y^{\rm int}-L_{\rm X}$}\\
Run      && $\alpha_{\rm T}$& $A_{\rm T} $ & $\alpha_{\rm L}$& $A_{\rm L} $\\
\hline
{\it Non-radiative} &   & 2.47 & 10.1 & 1.26 & 0.37\\
(bolometric) & & & & & \\
\hline
{\it Cooling}   & uncorrected   & 3.35 & 1.64 & 0.94 & 4.72\\
(soft-band)     & cooling-flow  &      &      &      &     \\
                & \hspace*{0.3cm} corrected     & 3.06 & 1.41 & 1.09 & 6.51\\
\hline
{\it Preheating}& uncorrected   & 3.26 & 1.93 & 0.98 & 6.65\\
(soft-band)     & cooling-flow  &      &      &      &     \\
                & \hspace*{0.3cm} corrected  & 3.18 & 1.70 & 1.04 & 8.09\\   
\hline
\end{tabular}
\end{table}

\subsection{The $Y-L_{\rm X}$ relation}

\begin{figure}
\centering \leavevmode\epsfysize=5.8cm \epsfbox{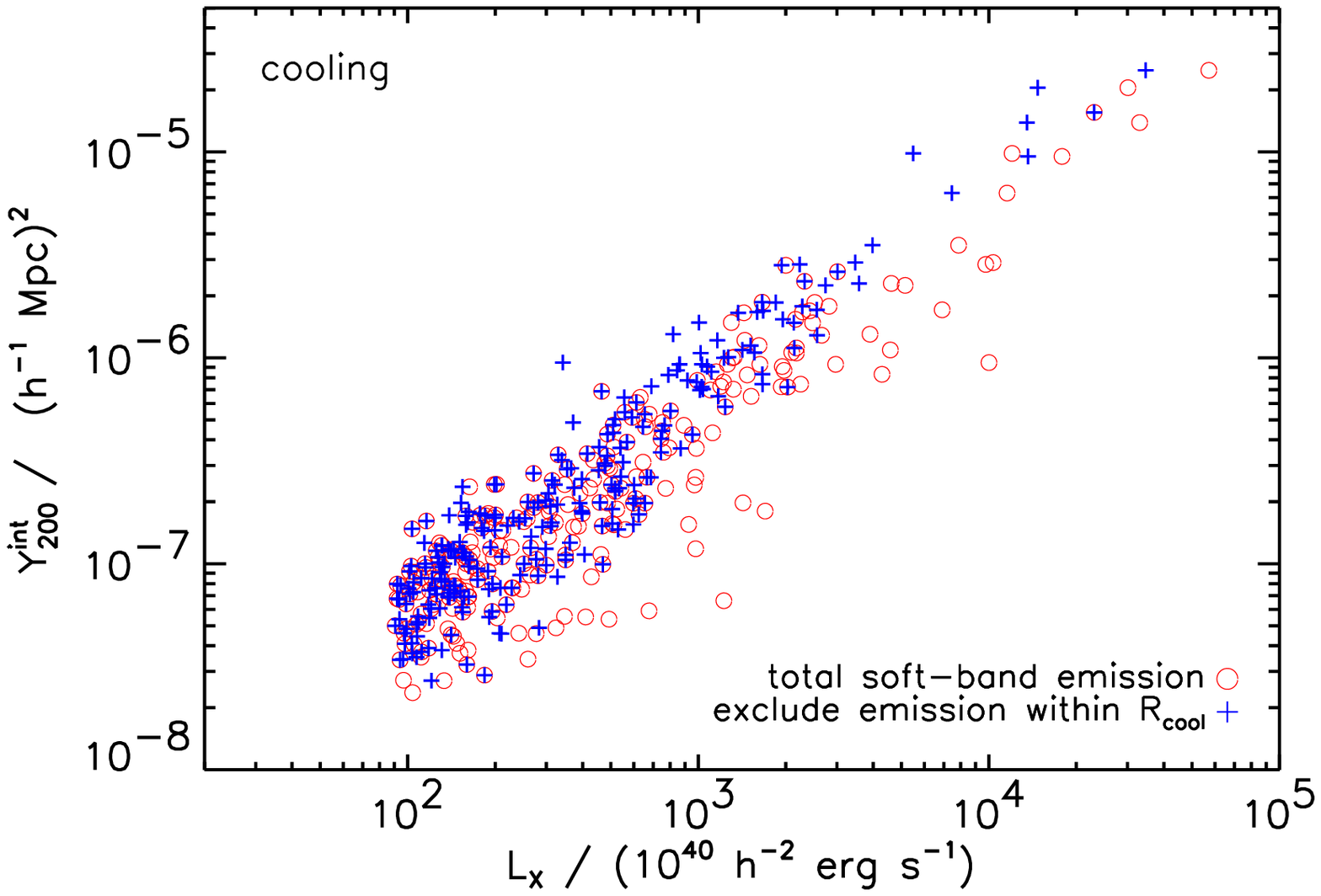}\\ 
\centering \leavevmode\epsfysize=5.8cm \epsfbox{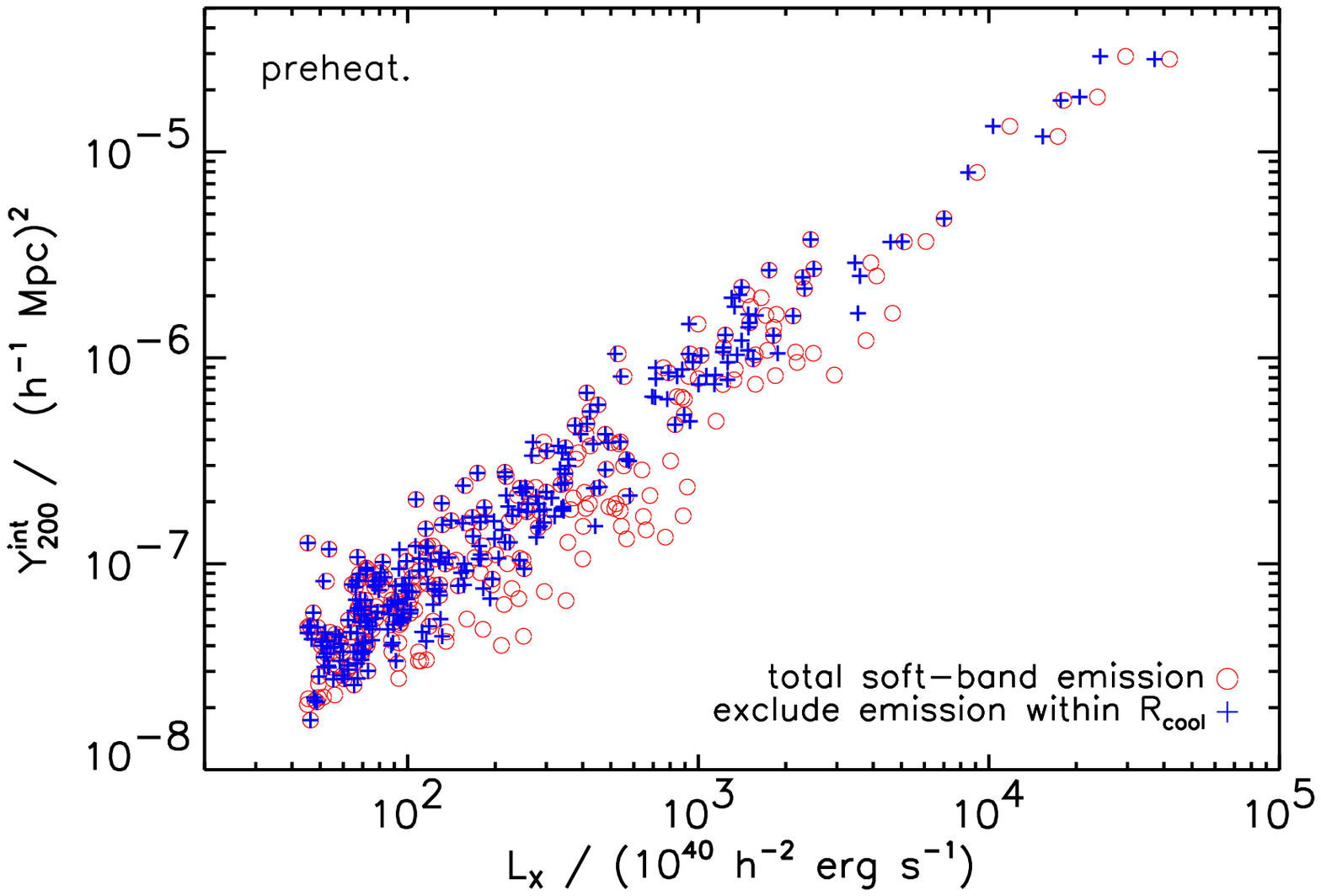}\\
\caption[ ]
{$Y^{\rm int}_{200}$ versus $L_{\rm X}$ for clusters at $z=0$ in the
{\it Cooling} (top panel) and {\it Preheating} (bottom panel) simulations.
Circles (crosses) denote objects when gas within the cooling
radius is included (excluded) in the calculations.}
\label{fig:lxb}
\end{figure}

We end this section by reporting on the $z=0$ correlation between 
$Y^{\rm int}$ and bolometric X-ray luminosity, $L_{\rm X}$, as inferred from 
the soft-band emission.
Fig.~\ref{fig:lxb} shows the $Y^{\rm int}_{200}-L_{\rm X}$ 
relation from the {\it Cooling} (top panel) and {\it Preheating} 
(bottom panel) simulations. Again, we use circles to represent values 
including all hot gas particles, whereas crosses represent results when
excluding gas from within the cooling radius.
The best-fit parameters $\alpha_{\rm L}$ and $A_{\rm L}$, defined by $Y^{\rm 
int}_{200}=(A_{\rm L}/10^7)\,(L_{\rm X}/L_{43})^{\alpha_{\rm L}}$,
where $L_{43}=10^{43}h^{-2}\, {\rm erg} \, {\rm s}^{-1}$,
are listed in Table~\ref{tab_1}. 

The {\it Cooling} and {\it Preheating} relations are significantly shallower 
than the predicted self-similar scaling, $Y \propto L^{5/4}$, which is well 
reproduced by the {\it Non-radiative} simulation 
(where $\alpha_{\rm L}=1.26$). 
Again, this is due to the increase in entropy of the gas, particularly
in low-mass clusters, which decreases their luminosity. Omitting
gas from within the cooling radius slightly increases the slope of the 
$Y^{\rm int}-L_{\rm X}$ relation and also reduces the scatter.

\section{Evolution of scaling relations}

We now investigate the evolution of the scaling relations
with redshift. In each of the following subsections, we factor out the 
dependence expected from self-similar evolution. Note that the self-similar 
evolution has no dependence on $\Omega_0$ because we define the overdensity 
with respect to the comoving critical density.

\subsection{Evolution of the $Y-M$ relation}

\begin{figure}
\centering \leavevmode\epsfxsize=8.1cm \epsfbox{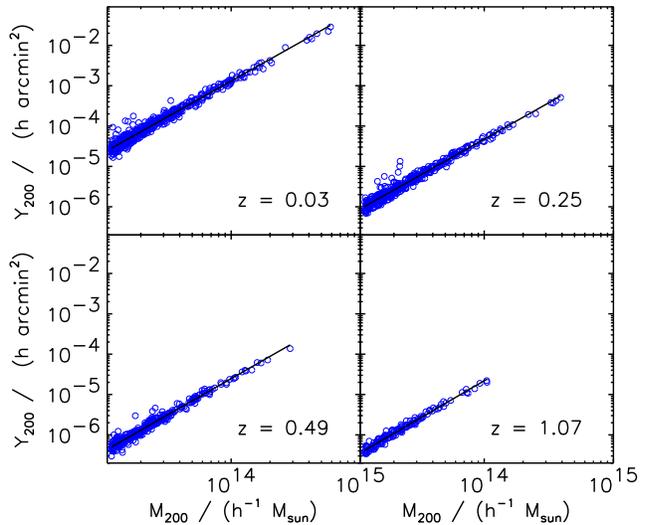}
\caption[ $Y_{200}-M_{200}$ relation for clusters in the
{\it Cooling}  simulation at various redshifts.]
{$Y_{200}-M_{200}$ relation at various redshifts for the {\it Cooling}  
simulation. The solid line is the power-law best fit.}
\label{fig:ymz} 
\end{figure}

Fig.~\ref{fig:ymz} shows the $Y_{200}-M_{200}$ relation for clusters
in the {\it Cooling} simulations, for a range of redshifts between
$z\sim 0$ and $z\sim 1$. 
Here we use  $Y_{200} = Y^{\rm int}_{200} d_{\rm A}^{-2}$, demonstrating the 
dominant effect of $d_{\rm A}$ on $Y$ at these redshifts.
Note that the number of clusters satisfying our mass
selection criteria, $M_{200}\gtrsim 1.18\times 10^{13} h^{-1}\,
{\rm M}_\odot$, decreases with increasing redshift.

\begin{figure}
\centering \leavevmode\epsfxsize=8.4cm \epsfbox{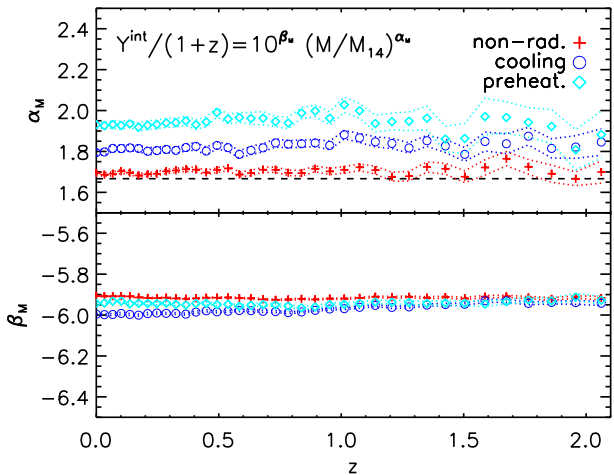}
\caption[]
{Redshift dependence of the slope ($\alpha_{\rm M}$) and normalization 
($\beta_{\rm M}$) of the cluster $Y^{\rm int}_{200}-M_{200}$ relation in the 
{\it Non-radiative} (crosses), {\it Cooling} (circles) and {\it Preheating} 
(diamonds) simulations. Symbols represent best-fit parameters and the 
dotted lines their 1-$\sigma$ uncertainties (see text). The dashed line
represents the slope predicted by the self-similar model 
($\alpha_{\rm M}=5/3$).
}
\label{fig:ymzalpha}
\end{figure}

A noticeable feature of Fig.~\ref{fig:ymz} is the apparent
independence of the slope of the $Y_{200}-M_{200}$ relation with
redshift. To quantify the effects of cooling and heating on the
evolution of the slope, we fit a power-law 
\begin{equation}
\frac{Y_{200}^{\rm int}}{1+z}=10^{\beta_{\rm M} (z)}\,
(M_{200}/10^{14}\,h^{-1}M_\odot)^{\alpha_{\rm M}(z)}\,,
\end{equation} 
to the distribution of clusters at each redshift, and plot 
the resulting best-fit values of $\alpha_{\rm M}$ against $z$ in the upper
panel of Fig.~\ref{fig:ymzalpha}. (We do not plot
results for redshifts beyond 2 as the number of clusters below our
mass threshold becomes too small for a reliable fit to be produced.)
Crosses, circles and diamonds represent best-fit values, and dotted lines 
denote 1-$\sigma$ uncertainties (estimated using the method given by 
Press et al.~1992) from the {\it Non-radiative}, {\it Cooling} and 
{\it Preheating} simulations respectively. 
Indeed, the results are consistent with an unevolving slope for all 
three models out to $z=2$.

The lower panel of Fig.~\ref{fig:ymzalpha} shows the evolution of the
normalization, obtained by fitting the same power-law function as
before, but with a fixed slope taken to be equal to the
mean value of $\alpha_{\rm M}(z)$ within the displayed range of $z$
(values are given in Table~\ref{tab_2}).
From this figure, we see that the evolution of the 
$Y_{200}^{{\rm int}}-M_{200}$ 
normalization is remarkably self-similar for all runs. Fitting a power-law,
$10^{\beta_{\rm M}(z)}\propto (1+z)^{\gamma_{\rm M}}$, to the derived 
normalizations (within the displayed range of $z$),
we find that the largest deviation from self-similar evolution of 
the normalization is only $\gamma_{\rm M}=0.14$, for the {\it Cooling} 
simulation (Table~\ref{tab_2}).

It is therefore clear that the evolution of the $Y_{\rm int}-M_{200}$ 
relation is not a sensitive probe of underlying non-gravitational 
physics, as it is a measure of the large-scale evolution of clusters.
As discussed in McCarthy et al.~(2003a), the effects of cooling and
preheating are much more apparent if the SZ flux density is evaluated
within a smaller radius, where the average gas density is higher.
As we shall see, such effects are apparent when studying evolution of
the X-ray properties of the clusters.

\begin{table}
\caption{Deviations from self-similarity in the mean slope and 
normalization of the SZ scaling relations. The quantities 
$\left< {\alpha_{\rm M}} \right>$, $\left< {\alpha_{\rm T}} \right>$ 
and $\left< {\alpha_{\rm L}} \right>$ are the mean values 
of the best-fit slopes of the $Y^{\rm int}_{200}-M_{\rm 200}$, 
$Y^{\rm int}_{200}-T_{\rm X}$ and $Y^{\rm int}_{200}-L_{\rm X}$ relations 
respectively, averaged over redshifts $z=0$ to $z=2$. 
The quantities $\gamma_{\rm M}$, $\gamma_{\rm T}$ and $\gamma_{\rm L}$ are 
power-law best fits to the normalization of the scalings, 
$10^{\beta(z)}\propto(1+z)^{\gamma}$; the first two are a good fit for the 
complete redshift range, but for luminosity the fit is restricted to $0<z<1$.
}
\label{tab_2}
\begin{tabular}{llccc}
\hline
         & &\multicolumn{3}{c}{Mean slope deviation from self-similarity} \\
Run      & &$\left< {\alpha_{\rm M}} \right>-5/3$ & $\left< {\alpha_{\rm T}} 
\right>-5/2$ & $\left< {\alpha_{\rm L}} \right>-5/4$ \\
\hline
{\it Cooling}       & & 0.16 & 0.44  & -0.16 \\
{\it Preheating}    & & 0.28 & 0.52  & -0.22 \\
\hline
         & & \multicolumn{3}{c}{Normalization deviation from self-similarity} \\
Run      & &$\gamma_{\rm M}$ & $\gamma_{\rm T}$ & $\gamma_{\rm L}$ \\
\hline
{\it Cooling}       & & 0.14 & 0.73 & 1.30 \\
{\it Preheating}    & & 0.02 & 0.33 & 1.62 \\
\hline
\end{tabular}
\end{table}

\subsection{Evolution of the $Y-T_{\rm X}$ relation}

\begin{figure}
\centering \leavevmode\epsfxsize=8.4cm \epsfbox{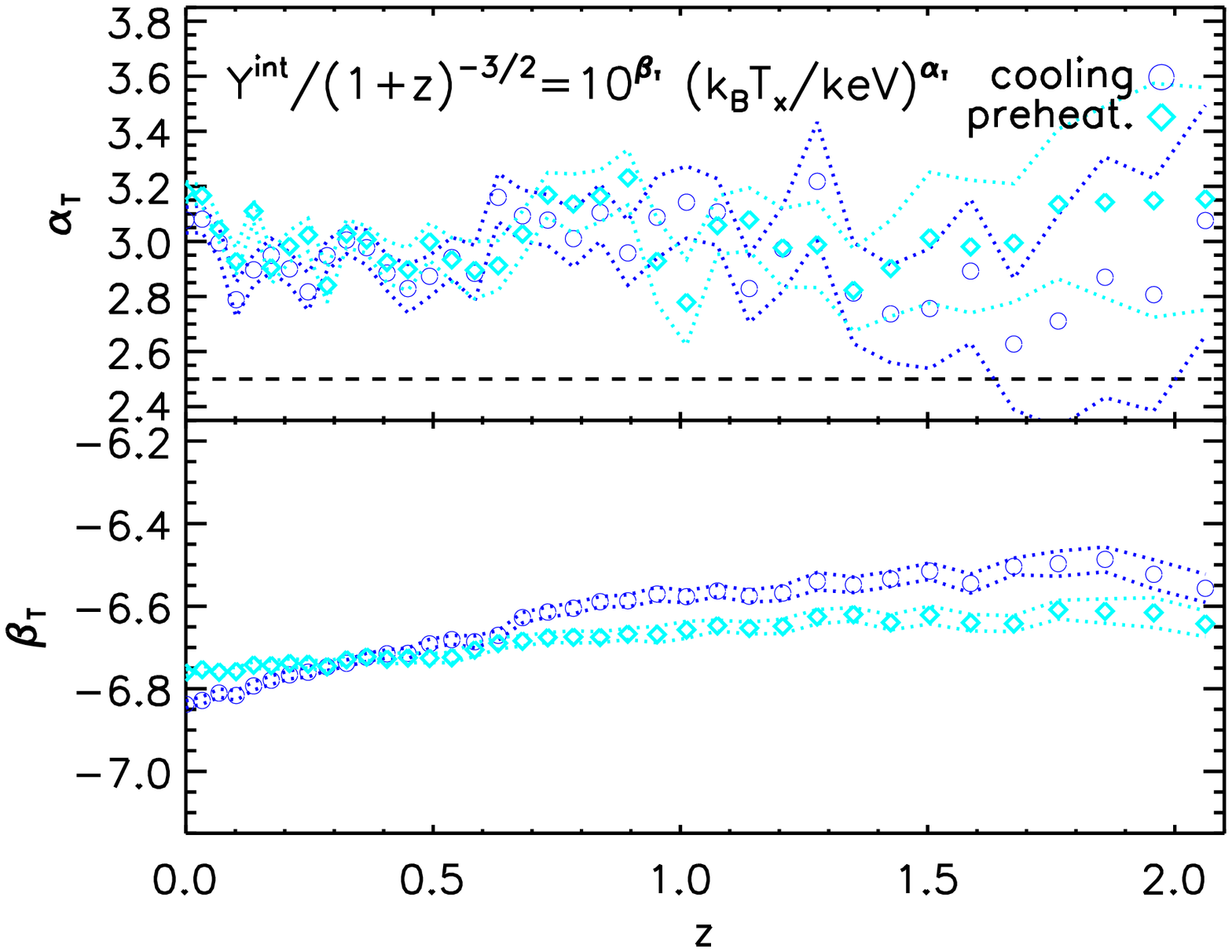}
\caption[]
{Redshift dependence of the slope ($\alpha_{\rm T}$) and normalization 
($\beta_{\rm T}$) 
of the cluster $Y^{\rm int}_{200}-T_{{\rm X}}$ relation in the 
{\it Cooling} (circles) and {\it Preheating} (diamonds)
simulations. Symbols represent best-fit values and dotted
lines their 1-$\sigma$ uncertainties. 
The dashed line represent the slope  predicted by the self-similar
model ($\alpha_{\rm T}=5/2$). 
}
\label{fig:ytxz}
\end{figure}

Fig.~\ref{fig:ytxz} shows the redshift dependence of the slope and
normalization of the $Y-T_{\rm X}$ relation for the {\it Cooling} and
{\it Preheating} simulations, using cooling-flow corrected temperatures.
As in the previous section, we first fit the cluster distribution at each 
redshift with a power-law 
\begin{equation}
\frac{Y_{200}^{\rm int}}{(1+z)^{-3/2}}=10^{\beta_{\rm T}(z)}\, 
(T_{\rm X}/{\rm keV})^{\alpha_{\rm T}(z)}\,,
\end{equation}
to determine the evolution of the slope. While this shows significant
statistical oscillations (reflecting the higher sensitivity of the
emission-weighted temperature to the thermal history of the gas),
there is no evidence of evolution with redshift beyond that predicted
by self-similarity. The slope is then fixed at the mean value, $\left<
\alpha_{\rm T}\right>$, before determining the redshift dependence of
the normalization, $\beta_{\rm T}(z)$. Unlike the $Y^{\rm int}$--$M$
relation, there is a definite evolutionary trend that is particularly
significant in the {\it Cooling} case. 
This is an indication that
evolution is predominantly in the X-ray properties rather then in the
SZ signal, which is much less sensitive to the details of the gas
distribution in the central regions of clusters.
Fitting the normalizations with
a power-law of the form $10^{\beta_{\rm T}(z)}\propto
(1+z)^{\gamma_{\rm T}}$, we find $\gamma_{\rm T}= 0.73$ and
$\gamma_{\rm T}= 0.33$ for the {\it Cooling} and {\it Preheating}
simulations, respectively. In both cases the normalization of the
$T_{{\rm X}}$--$M$ relation increases with time compared to the
expected self-similar evolution.

\subsection{Evolution of the $Y-L_{\rm X}$ relation}

We end by studying the evolution of the 
$Y-L_{\rm X}$ relation, shown in Fig.~\ref{fig:ylxz}.
Our power-law fitting function was 
\begin{equation}
\frac{Y_{200}^{\rm int}}{(1+z)^{-27/8}}=10^{\beta_{\rm L}(z)}\,
(L_{\rm X}/L_{43})^{\alpha_{\rm L}(z)}\,,
\end{equation} 
using cooling-flow corrected values for $L_{\rm X}$.
Again, the slopes are constant with redshift to good
approximation, and there is additional evolution in the normalization. In this 
case a single power-law is not so good a fit over the complete redshift range, 
and we restrict the fit to $0<z<1$ which is where observations are likely 
to be made. Fitting a power-law
$10^{\beta_{\rm L}(z)}\propto(1+z)^{\gamma_{\rm L}}$
yields $\gamma_{\rm L} \sim 1.3$ \& $1.6$ in the {\it Cooling} and
{\it Preheating} cases respectively (Table~\ref{tab_2}) in this redshift 
range, with the relation flattening at redshifts beyond one.

\section{Conclusions}

In this paper, we investigated the relationship between
SZ and X-ray properties of clusters drawn from three
simulations: a {\it Non-radiative} simulation, where the gas could
only heat adiabatically and through shocks; a {\it Cooling} simulation,
where the gas could also cool radiatively and a {\it Preheating} simulation,
where we additionally heated the gas by 1.5 keV per particle at $z=4$.
We focussed on the relations between the thermal SZ signal $Y_{{\rm int}}$ and 
cluster mass,
temperature and X-ray luminosity, both at $z=0$ and 
the evolution with redshift. Our main conclusions are as 
follows:

\begin{figure}
\centering \leavevmode\epsfxsize=8.4cm \epsfbox{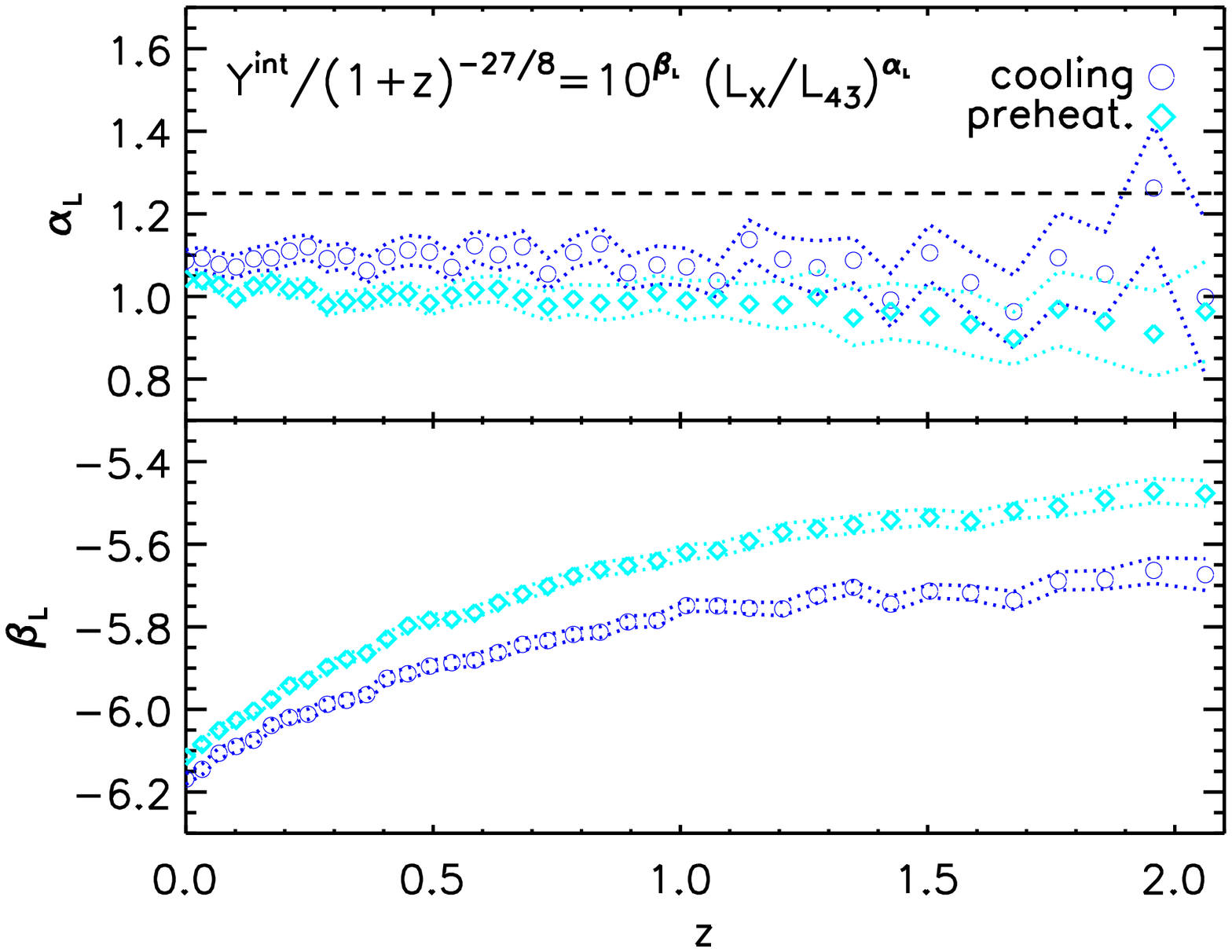}
\caption[]
{Redshift dependence of the slope ($\alpha_{\rm L}$) and normalization 
($\beta_{\rm L}$) of the cluster $Y^{\rm int}_{200}$--$L_{{\rm X}}$ relation 
in the {\it Cooling} (circles) and {\it Preheating} (diamonds) simulations. 
Symbols represent best-fit parameters and dotted lines their 
1-$\sigma$ uncertainties. 
The dashed line represent the slope predicted by the self-similar 
model ($\alpha_{\rm L}=5/4$).
}
\label{fig:ylxz}
\end{figure}

\begin{itemize}

\item $Y_{{\rm int}}$ is tightly correlated with both mass
and mass-weighted temperature. Scaling relations of $Y_{{\rm int}}$
with X-ray quantities show significantly more scatter however,
due to the sensitivity of the latter (which are emission-weighted)
to the gas distribution in the cores of clusters. Excluding gas
from within the cooling radius of each cluster reduces the scatter. 

\item The {\it Non-radiative} simulation scaling relations closely
match those predicted by the self-similar model, with \mbox{$Y_{{\rm int}} 
\propto
T_{\rm X}^{5/2}$} and $Y_{{\rm int}} \propto L_{\rm X}^{5/4}$. 

\item {\it Cooling} decreases $Y_{{\rm int}}$ for a given cluster mass, 
due to the reduction in the hot gas fraction (which decreases 
$Y_{{\rm int}}$) being more effective than the increase in temperature 
of the remaining gas (which increases $Y_{{\rm int}}$).  
{\it Preheating} the gas has a similar effect except in the most massive 
clusters, where the heating energy was not sufficient to expel much gas 
from the halo. At redshift zero, the resulting $Y_{{\rm int}}$--$T_{\rm X}$ 
scaling relation is steeper than the self-similar relation, 
$Y_{{\rm int}} \propto T_{\rm X}^{3.0-3.4}$, and the 
$Y_{{\rm int}}$--$L_{\rm X}$
relation flatter,  $Y_{{\rm int}} \propto L_{\rm X}^{0.9-1.1}$.

\item There is no evidence of evolution in the slopes of any of the scaling
relations.  The normalization of the $Y_{200}-M_{200}$ relation is also 
consistent with self-similar evolution.  However, the relations between 
$Y_{{\rm int}}$ and the cluster X-ray properties do show significant 
evolution in normalization relative to that expected from self-similarity, 
indicating that evolution is predominantly in the X-ray properties rather 
than the SZ properties. The effects of excess entropy reduces the negative 
evolution of the $Y_{{\rm int}}$--$T_{\rm X}$ and $Y_{{\rm int}}$--$L_{\rm X}$
with redshift, particularly at $z<1$.

\end{itemize}

\section*{Acknowledgments}
A.C.S.~was supported by the EU CMBNET network, S.T.K.~by PPARC, and A.R.L.~in 
part by the Leverhulme Trust. The simulations used in this paper were carried 
out on the Cray-T3E (RIP)
at the EPCC as part of the Virgo Consortium programme of
investigations into the formation of structure in the Universe. 
A.C.S.~acknowledges the use of the computer facilities at CALMIP-CICT, 
France, and visits to Sussex supported by PPARC.


\end{document}